# Nonreciprocal resonant transmission/reflection based on a one-dimensional photonic crystal adjacent to the magneto-optical metal film

Cheng He[1,2], Chang-Sheng Yuan[1], Ming-Hui Lu[1], Yan-Feng Chen[1*], Cheng Sun[2]

[1]National Laboratory of Solid State Microstructures & Department of Materials Science and Engineering, Nanjing University, Nanjing 210093, People's Republic of China

[2]Department of Mechanical Engineering, Northwestern University, Evanston, IL 60208-3111, USA

We report the design of nonreciprocal resonant transmission/reflection using a one-dimensional photonic crystal (1DPC) adjacent to the magneto-optical (MO) metal film. The nonreciprocal surface modes are found at the interface between the PC and MO metal within the forbidden band of native PC structure. Breaking time-reversal symmetry using external magnetic field gives rise of such unique nonreciprocal properties. Quantitatively understanding of nonreciprocal resonant optical transmission/reflection behavior is performed using an effective admittance-matching theory. With excitation of the nonreciprocal surface modes, light can transmit and be reflected in one-way. Such design offers promising potential in realizing the optical diode.

[*] e-mail: yfchen@nju.edu.cn

The topic of one-way wave propagation has attracted much interest in both optics and acoustics. Several ways have been proposed, such as nonreciprocal linear, nonlinear, phase shift and diffractive process[1-10]. Symmetry-breaking plays a crucial role in realizing these one-way phenomena, no matter which mechanism is used. To break Time-reversal (T) symmetry, it is a general way to introduce MO medium in the presence of an external magnetic field. For example, various types of Giga-Hertz one-way electromagnetic surface modes in MO dielectric PCs have been both theoretically and experimentally studied[3,4,11-15]. Surface plasmon polaritons (SPPs) of MO metal can also exhibit nonreciprocal behaviors[16]. The one-way waveguide modes at the interface of the two-dimensional dielectric PC and MO metal has been proposed due to the different cutoff frequency between forward and backward SPPs[17]. After applying external magnetic field on the MO metal to break T symmetry, nonreciprocal SPP waves could be excited to realize the one-way propagation.

On the other hand, dielectric-metal and metal metamaterials have shown their abilities in manipulating evanescent waves to achieve super resolution[18,19]. Evanescent waves, characterized by the imaginary propagating constant without carrying energy, play a crucial role in these models, which could be enhanced in negative refractive index medium or coupled to propagating mode using artificial structures. Surface modes, with imaginary propagating constant perpendicular to the boundary, could perform similar coupling effects as that of evanescent waves, such as extraordinary optical resonant transmission[20-23]. However, it is unclear whether the nonreciprocal surface modes can be enhanced or coupled into free space, and this coupling mechanism still needs to be studied.

In this Letter, focusing on exciting, enhancing and coupling the nonreciprocal surface modes,

we presented a type of nonreciprocal resonant transmission/reflection model based on 1DPC-metal structure. An effective admittance-matching theory based on such nonreciprocal model was proposed to determine the operating frequency quantitatively. With external magnetic field, the forward and backward admittances of MO metal become different to excite nonreciprocal SPP modes, which can be enhanced and coupled into free space, realizing nonreciprocal light propagation in the band-gap of native PC. This model is useful to realize the optical diode in experiment for its advantage in the low loss, small area of an external magnetic field, wide operating frequency window, and compatible technique in industry.

Herein, we constructed a 1DPC-metal structure with 8 periods symmetric $TiO_2/SiO_2/TiO_2$ PC attached a thin MO metal film as shown in Fig. 1(a). The lattice constant of PC is $0.6\lambda_p$ with thickness of each layer $0.2\lambda_p$, and the thickness of MO metal film is $0.4\lambda_p$, where $\lambda_p$ is the bulk plasmon wavelength. The refractive indices of $SiO_2$ and $TiO_2$ are 1.46 and 2.25 respectively. By applying external magnetic field **B** in *z* direction (covering the thin metal film), the dispersive dielectric function of MO metal yields

$$\vec{\varepsilon}(\omega) = 1 - \frac{\omega_p^2}{(\omega+i/\tau)^2 - \omega_b^2} \begin{bmatrix} 1+i\frac{1}{\tau\omega} & i\frac{\omega_b}{\omega} & 0 \\ -i\frac{\omega_b}{\omega} & 1+i\frac{1}{\tau\omega} & 0 \\ 0 & 0 & \frac{(\omega+i/\tau)^2 - \omega_b^2}{\omega(\omega+i/\tau)} \end{bmatrix}, \qquad (1)$$

where $\omega_p$ is the bulk plasmon frequency, $\omega_b = eB/m$ represents the cyclotron frequency, and *e* is the electron charge and *m* is the electron mass. Metal loss is indicated by decay time $\tau$. To calculate the optical admittance (the ratio of the total magnetic to electric **H/E** field) of this nonreciprocal metal, we assume the loss to be infinitesimal. For TE mode (the electric field along *z* direction), the optical admittance of MO metal can be written as

$$\eta_m^{\pm} = -\eta_0 \frac{n^2 \omega \varepsilon_d}{i\varepsilon_f c k_y - \varepsilon_d \sqrt{n^2 \omega^2 - c^2 k_y^2}}, \qquad (2)$$

where $\varepsilon_f = -\omega_p^2 \omega_b/(\omega^3 - \omega \omega_b^2)$, $\varepsilon_d = 1 - \omega_p^2/(\omega^2 - \omega_b^2)$, $n = \sqrt{(\varepsilon_d^2 - \varepsilon_f^2)/\varepsilon_d}$, $c$ is the speed of light in vacuum, $\eta_0$ is the admittance of vacuum, and $k_y$ represents the wave vector along the y direction. The optical admittance of metal is reciprocal $\eta^+ = \eta^-$ ($\varepsilon_f = 0$) without the external magnetic field. By applying the external magnetic field, T symmetry is broken. Then for a pair of counter incident waves, where $+k_y$ and $-k_y$ have opposite signs in Eq. (2), the admittance of two cases are different, which would excite the nonreciprocal surface modes and make nonreciprocal transmission/reflection occur. It should be noticed that for normal incident wave ($k_y = 0$), the nonreciprocal phenomena would not occur even with external applied magnetic field. The difference of admittance between two counter incident waves could be enlarged by increasing the angle of incident wave or the intensity of external applied magnetic field.

On the other hand, the optical admittance of such three layers symmetric PC as in Fig. 1(a) has been well-discussed[24-27]. The total characteristic matrix can be represented mathematically by a single equivalent characteristic matrix[28]. The optical admittance of PC for TE mode is

$$\eta_{PC} = \frac{\eta_p [\sin(2\delta_p)\cos(\delta_q) + \rho^+ \cos(2\delta_p)\sin(\delta_q) - \rho^- \sin(\delta_q)]}{\sin\{a c \cos[\cos(2\delta_p)\cos(\delta_q) - \rho^+ \sin(2\delta_p)\sin(\delta_q)]\}}, \qquad (3)$$

where subscript $p$ and $q$ represents the TiO$_2$ and SiO$_2$ layer respectively. $\eta_{p,q} = \eta_0 n_{p,q}^2 \omega / k_{x(p,q)}$ is the optical admittance of a single layer, $\delta_{p,q} = k_{x(p,q)} d_{p,q}$ is the optical phase of a single layer, and $\rho^{\pm} = (\eta_p/\eta_q \pm \eta_q/\eta_p)/2$. $k_{x(p,q)} = \sqrt{n_{p,q}^2 \omega^2 - k_{y(p,q)}^2}$ is the wave vector along $x$ direction, $d$ is the thickness, and $n$ is the index of refraction.

The condition of existence of the surface mode at the interface of metal-1DPC is

$$\eta_{PC} = -\eta_m. \qquad (4)$$

Since the nonreciprocal character of optical admittance of MO metal in Eq. (2), surface modes at the interface of metal-1DPC with a pair of counter incident waves would be nonreciprocal. These surface modes would locate in the band-gap of native PC, where both optical admittance of PC and metal are pure imaginary. So we just need to analyze the imaginary part of optical admittance. In Fig. 1(b), we show the frequency dependent imaginary part of optical admittance of the MO metal and PC with a pair of $\pm 30^o$ incident waves. The nonzero region of the imaginary optical admittance of PC indicates the first band-gap from $0.3854\omega_p$ to $0.4755\omega_p$. The imaginary optical admittances of MO metal correspond to forward (+$y$ direction) and backward (-$y$ direction) excitation of surface modes, where the cyclotron frequency (induced by external magnetic field) is assumed to be $\omega_b = 0.1\omega_p$. The different frequencies of intersection $0.4269\omega_p$ and $0.4227\omega_p$ correspond to the backward and forward surface modes, respectively.

With external applied magnetic field, optical admittances for forward and backward waves would be split. The effective permittivity of metal is negative in optical window, and the effective index of refraction is imaginary. For the interface of MO metal-air, the nonreciprocal SPP along the interface could be excited as reported before[17]. To couple this nonreciprocal SPP and make it radiate into free space, another effective single negative material is needed. The region of the band-gap of 1DPC could be chosen as an excellent candidate, since the suitable operating frequency window and compatible technique in industry. The field of incident wave would be intensely amplified at the interface of 1DPC-metal, which would be excited by coupling and resonance of waves into free space. Furthermore, some other models might also realize the similar process, such as metamaterial-metal and structured MO metal models.

Then, we calculate the surface modes at the interface of MO metal-1DPC in the first band-gap according to Eq. (4) [Fig. 2(a)], where $\omega_b = 0.1\omega_p$. According to the boundary condition of Maxwell equations, the incident angle exciting the surface wave can be defined as $\theta = \arcsin[(k/(\omega/c)]$, where $k$ and $\omega$ correspond to the wave vector and frequency of surface mode as in Fig. 2(a). Fixing an incident angle, the backward and forward surface modes would be excited at different frequencies to realize the one-way properties. With normal incident wave, the forward and backward surface modes locate at the same frequency. Increasing the incident angle could make the nonreciprocal frequency blue shift. The surface modes above the air line can be excited by incident waves from air. To excite the surface modes below the air line, some high index of refraction medium or structured surface should be needed to obtain the large $k_y$. To verify the admittance-matching theory for these nonreciprocal surface modes, we use various methods to get the same results shown in Fig. 2(b). The solid lines are calculated by theoretical Eq. (4). The open circles are calculated by supercell method, constructed by air-1DPC-metal-air model, where the thickness of each air layers is $3\lambda_p$. The pluses are obtained by extracting the peak values from transmission spectra. A little differences come from finite layers 1DPC and metal and the influence of air layers[23]. Then, we plot the nonreciprocal transmission spectra in Fig. 3(a) with $\omega_b = 0.1\omega_p$. Three incident cases $\pm 15^o$ (dashed), $\pm 30^o$ (solid line) and $\pm 45^o$ (open circles) are calculated. The maximum transmission almost reaches one. The relative nonreciprocal frequency ($\Delta\omega/\omega$) is about 15%. The maximum value of contrast ratio (($T^+ - T^-)/(T^+ + T^-)$, $T^\pm$ is transmission) is about 0.8 [Fig. 3(b)]. Both the values and shapes of transmission can be manipulated by changing the thickness of the interface layer $TiO_2$ between 1DPC and MO metal[23,25]. Moreover, the contrast ratio could be enlarged by increasing the external magnetic field. As shown in Figs. 3(c) and 3(d), for $\pm 30^o$ incident cases, if the external applied magnetic field is $\omega_b = 0.3\omega_p$, the relative nonreciprocal frequency would be over 30% [solid line in Fig. 3(c)], and the contrast ratio is over 0.95 [solid line in Fig. 3(d)]. Due to the thin thickness of metal, the wave attenuation distance in the metal is short. This 1DPC-metal model has a good merit of the low loss. Considering a typical loss of metal, $1/\tau = 5 \times 10^{-3} \omega_p$, the transmission can also reach 70% [open circles in Fig. 3(c)].

The field distributions with $\pm 30^o$ incident waves are shown in Fig. 4 [open circles in Fig. 3(c)]. Only the forward surface mode could be excited at frequency $0.4175\omega_p$. So only the incident waves from lower part (left-down and right-down) with the wave vector along +y direction ($+k_y$ component), could transmit as shown in Fig 4(a). The electromagnetic field is intensely amplified to propagate toward to the interface of metal-PC, and then transmit through the boundary and attenuate, at last transmit into the air. A pair of counter incident electromagnetic waves would have different propagating phenomena: one is almost completely transmit and the other is totally reflected[29]. The one-way property would be reversed at frequency $0.4293\omega_p$ as shown in Fig. 4(b). Only the incident waves from upper part (left-up and right-up) with the $-k_y$ component could excite the backward surface modes and transmit. Furthermore, these two cases of one-way propagation would be reversed by reversing the external applied magnetic field.

The nonreciprocal effects are attributed to the T symmetry broken under the external magnetic field. On the other hand, spatial inversion symmetry in real space is relative to x direction, and in reciprocal space the parity is relative to $k_y$ direction. That is to say that a pair of incident waves symmetry relative to y direction is the parity-time symmetric case, while a pair of counter incident waves is one-way transmission and a pair of symmetric incident waves relative to x direction is one-way reflection[29].

In summary, we have designed a 1DPC-MO metal model to realize the nonreciprocal resonant transmission/reflection, which stems from the excitation of nonreciprocal surface modes at the interface between the PC and metal, in the band-gap of pure PC with the broken time-reversal T symmetry applying an external magnetic field. An effective nonreciprocal admittance-matching theory is proposed. The advantage of this 1DPC-metal model with of the low loss (propagate through the thin metal film), small area with an external magnetic field (just need to cover the thin metal film), wide operating frequency window (dependent on the incident angle), and compatible technique in industry, may be very useful to design some optical nonreciprocal devices.

The work was jointly supported by the National Basic Research Program of China (Grant No. 2012CB921503) and the National Nature Science Foundation of China (Grant No. 51032003). We also acknowledge the support from the Nature Science Foundation of Jiangsu Province (Grant No. BK2009007), Academic Program Development of Jiangsu Higher Education (PAPD), and China Postdoctoral Science Foundation (Grant No. 2012M511249).

**Figure Captions**

FIG. 1 (Color online). (a) Schematic of 1DPC- metal model including 8 periodical $TiO_2/SiO_2/TiO_2$ PC and MO metal film. (b) The imaginary part of optical admittances of the PC $(-\eta_{PC})$ and MO metal $(-\eta_m^{\pm})$ with a pair of $\pm 30^o$ incident waves (external magnetic field $\omega_b = 0.1\omega_p$). The dot-dashed line is the optical admittance of metal $(-\eta_m^0)$ without external magnetic field.

FIG. 2 (Color online). (a) The nonreciprocal surface modes at the interface of MO metal-1DPC of the first band-gap ($\omega_b = 0.1\omega_p$). The shadow regions are the projected band of 1DPC along $k_y$ direction. (b) Zoom-in surface modes are calculated by various methods: the solid lines by theoretical Eq. (4), the open circles by supercell method and the pluses by transmission spectra.

FIG. 3 (Color online). (a) The nonreciprocal transmission spectra and (b) contrast ratio of backward and forward waves with incident angles $\pm 15^o$ (dashed line), $\pm 30^o$ (solid line), and $\pm 45^o$ (open circles), the external magnetic field $\omega_b = 0.1\omega_p$. (c) The transmission spectra and (d) contrast ratio of $\pm 30^o$ incident waves, where dashed ($\omega_b = 0.1\omega_p$) and solid ($\omega_b = 0.3\omega_p$) lines represent lossless condition, open circles represent $\omega_b = 0.3\omega_p$ and $1/\tau = 5\times 10^{-3}\omega_p$.

FIG. 4 (Color online). The field distributions with $\pm 30^o$ incident waves [open circles in Fig. 3(c)]. (a) Exciting the forward surface mode at frequency $0.4175\omega_p$. (b) Exciting the backward surface mode at frequency $0.4293\omega_p$.

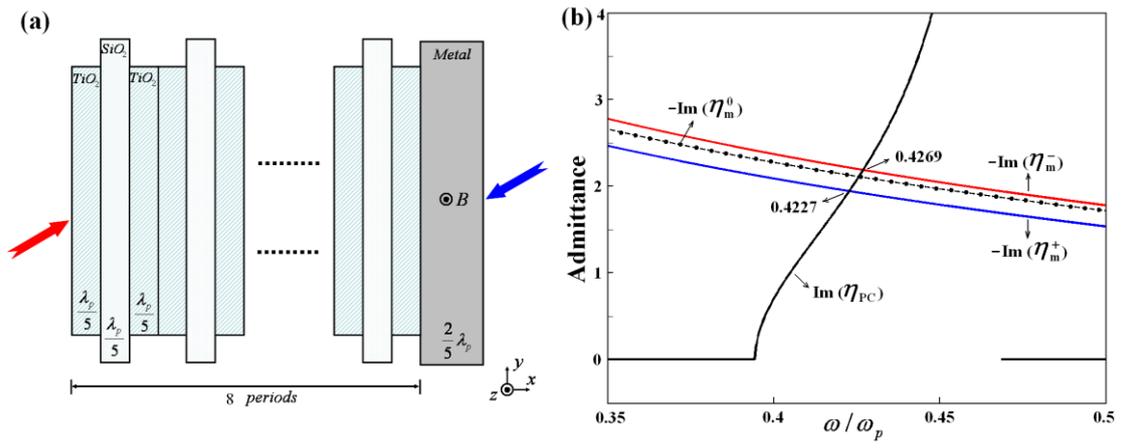

Figure 1

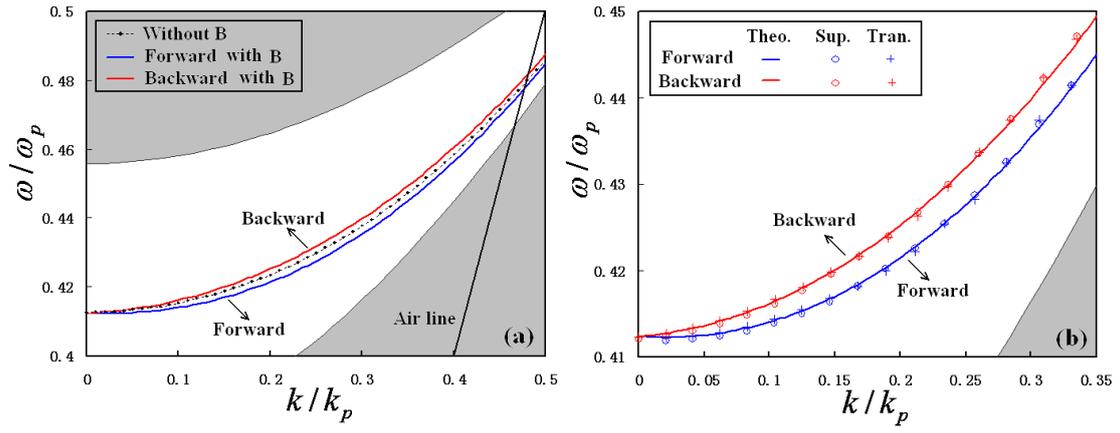

Figure 2

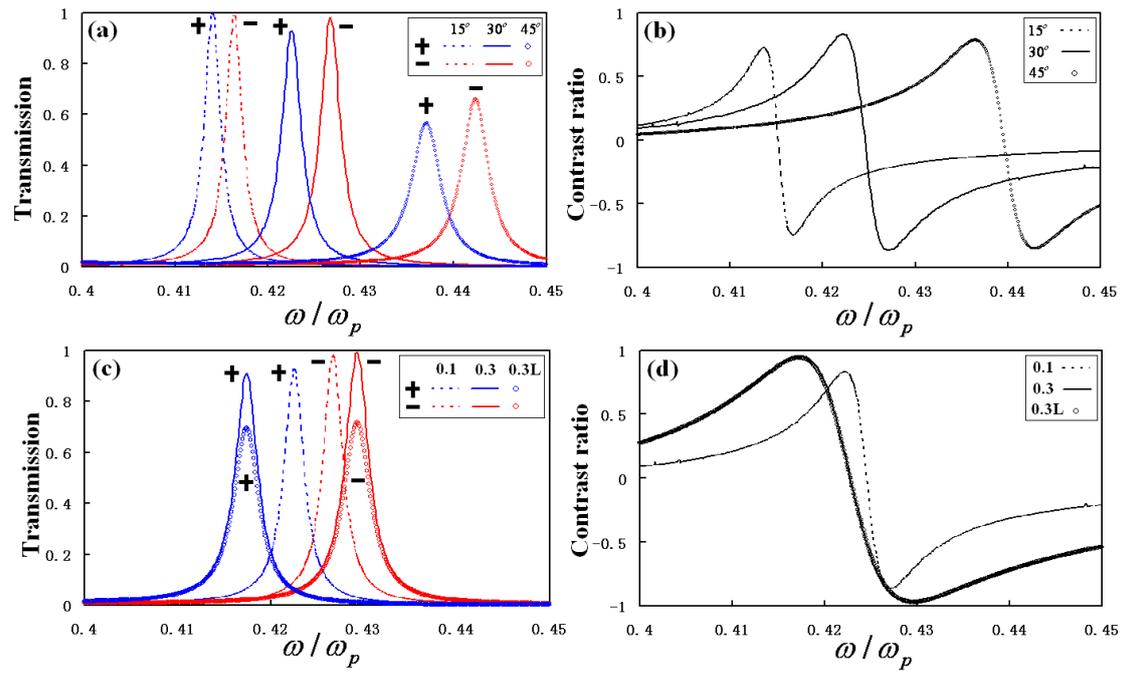

Figure 3

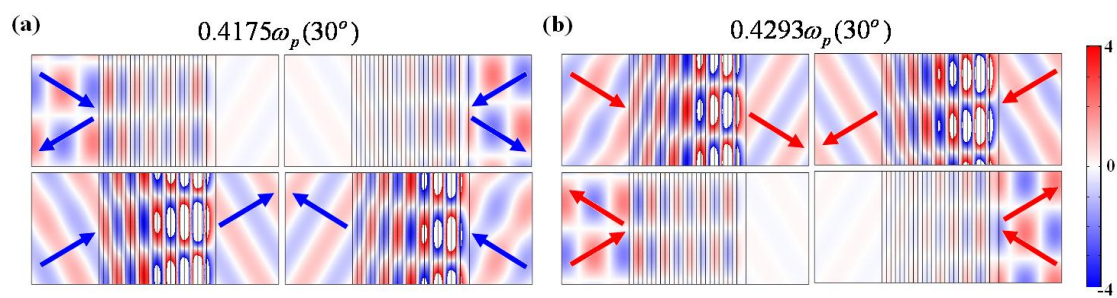

Figure 4